# Response and Feedback of Cloud Diurnal Cycle to Rising Temperatures

Jun Yin, Amilcare Porporato*

Department of Civil and Environmental Engineering, Duke University, Durham, North Carolina, USA.

*Correspondence to: amilcare.porporato@duke.edu

**By reflecting solar radiation and reducing longwave emissions, clouds regulate the earth's radiation budget, impacting atmospheric circulation and cloud dynamics[1]. Given the diurnal fluctuation of shortwave and longwave radiation, a shift in the cloud cycle phase (CCP) may lead to substantial feedbacks to the climate system[2]. While most efforts have focused on the overall cloud feedback[3,4], the response of CCP to climate change has received much less attention. Here we analyze the variations of CCP using both long-term global satellite records and general circulation models (GCM) simulations to evaluate their impacts on the earth's energy budget. Satellite records show that in warm periods the CCP shifts earlier in the morning over the oceans and later in the afternoon over the land. Although less marked and with large inter-model spread, similar shifting patterns also occur in GCMs over the oceans with non-negligible CCP feedbacks. Over the land, where the GCM results are not conclusive, our findings are supported by atmospheric boundary layer models. A simplified radiative model further suggests that such shifts may in turn cause reduced reflection of solar radiation, thus inducing a positive feedback on climate. The crucial role of the cloud cycle calls for increased attention to the temporal evolution of the cloud diurnal cycle in climate models.**

Cloud feedback is argued to be one of the largest sources of uncertainty in climate prediction[5,6]. Because of its significant impact on the earth's radiation budget, its quantification has attracted considerable attention in the recent literature[1,7]. Clouds shade the solar radiation during daytime, but also have greenhouse effects, largely independent of their timing, which reduce the emission of longwave radiation to the outer space[8]. It is thus logical to expect that even small differences in the timing of the maximum cloud amount (hereafter referred to as

'cloud cycle phase' or CCP) could have large feedback to the earth's climate system[2,9]. Previous studies on cloud feedbacks have focused more on the change of mean cloud amount[10,11], cloud levels, and other properties[3], without paying enough attention to the CCP. Cloud radiative forcing approach[12] have been successful in evaluating the overall cloud feedback at sub-daily scale[13], but only represent integrated effects due to the changes of all types of cloud properties[14]. As a starting point for assessing the specific role of the cloud diurnal cycle, in this work we quantify the CCP shift from cold to warm periods using the International Satellite Cloud Climatology Project (ISCCP) satellite records[15] and the outputs of nine General Circulation Models (GCMs). These shifts are then linked to the net radiation changes at the tropopause by using a simplified radiation balance model. Finally, the radiation changes and their geographical distributions are analyzed to evaluate the importance of shifts in CCP.

We begin by decomposing the cloud diurnal cycle, characterized by its variation in cloud fraction[9,13], into Fourier modes (see Methods). We focus on three terms: the mean ($f_m$), the amplitude ($f_a$), and the phase ($t_c$) of the first harmonic, which account for the primary characteristics of cloud diurnal variation[2]. When applied to the 3-hour ISCCP D1 satellite data set, such quantities show contrasting land-ocean patterns and seasonal variations (see Figure 1 and Figure S1). Regions of high cloud fractions over the tropics mark the locations of Intertropical Convergence Zone (ITCZ) and its northward swing in summer brings heavy rainfall to India and southeast China. Clouds over the land show larger diurnal variation with maximum amount in the late afternoon, while over the oceans they have smaller diurnal amplitudes and early morning peaks, consistent with several other studies[9,16,17]. Warm seasons (boreal summer in north hemisphere and austral summer in south hemisphere) usually have more cloud coverage, later phase, and larger amplitude.

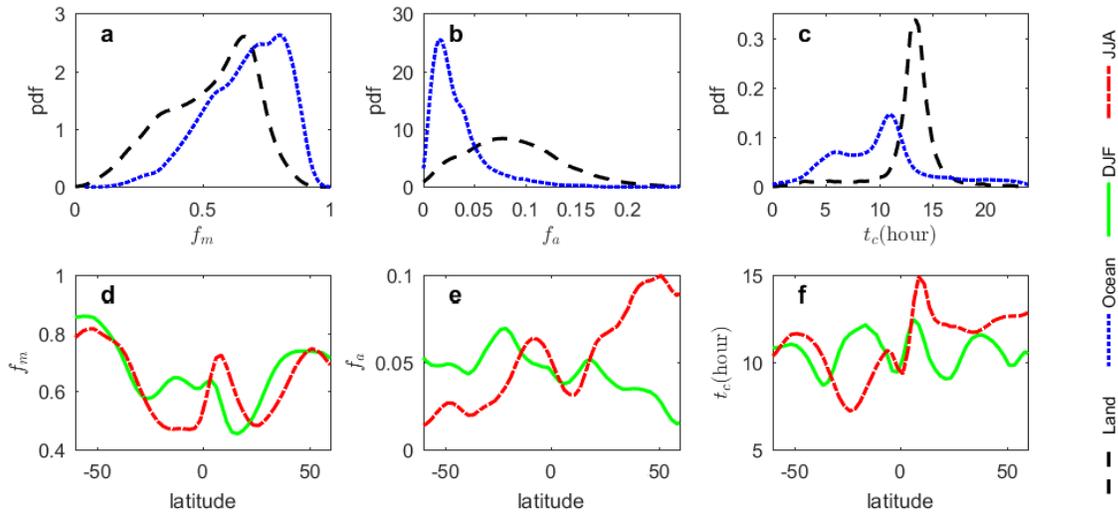

**Figure 1 | Statistics of cloud diurnal cycle.** The empirical probability distribution function (pdf) of (**a**) mean, (**b**) amplitude, and (**c**) phase of cloud diurnal cycle over (black dash) land and (blue dot) ocean within 60S-60N of latitude; zonal average of (**d**) mean, (**e**) amplitude, and (**f**) phase of cloud diurnal cycle in boreal summer (JJA: red dash-dot) and winter (DJF: green solid). The cloud diurnal cycle climatology is derived from ISCCP D1 3-hour data sets from 1984 to 2009.

The CCP discussed above is driven by multiple factors such as the diurnal patterns of solar energy absorption, large-scale subsidence, sea/land breeze, and boundary-layer dynamics[9,16]. These factors are projected to have different changes in response to the global warming[1,18], possibly resulting in CCP shift with increasing mean surface temperature. To test this hypothesis, we again used the ISCCP D1 data sets and analyzed the shift in CCP ($\Delta t_c$) between two periods (1984-1996 and 1997-2009) characterized by an average surface temperature difference of 0.27 K, according to the Goddard Institute for Space Studies (GISS) analysis of global surface temperature change[19]. The ISCCP has been reported to suffer from artifacts that influence the analysis of long-term cloud trends due to the slowly varying systematic bias[20,21]. This systematic bias is expected to propagate into the mean and amplitude values. Not surprisingly, in fact, we found that there are strong (and fake) decreasing trends in both $f_m$ and $f_a$. However, the estimation of the CCP is expected to be less impacted by the artifacts because it only involves the observation at hourly timescale, which is quite different from the systematic bias over the timescale of decades. The results of the shift in CCP going from the colder period 1984-1996 to the warmer period 1997-2009 are shown in Figure S2. In general, the clouds in the warm period were found to peak a little earlier over the ocean and slightly later over the land. In order to minimize the potential effects of artifacts in the satellite data, we checked for strong correlation between CCP and surface temperature, based on the fact that such a correlation, while less affected by the systematic biases in the satellite record, is a necessary condition for feedback[22]. Removing regions with weak correlations between CCP and earth surface temperature time series (see Figure S3) made the shifting patterns more distinctive and allowed us to highlight locations with more realistic CCP feedbacks within the ISCCP records.

In order to understand the potential impacts of these trends, we first considered the daily average of the so-called radiative kernel with respect to the CCP, $\partial \overline{R} / \partial t_c$, where $\overline{R}$ is the daily-average net radiation at the tropopause (see Methods). In a nutshell, a radiative kernel is the changing rate of the earth's radiation budget due to the variation of a specified climate variable (in this case $t_c$) with the others kept constant[7,11]. In principle, the cloud radiative kernels could be computed from climate models. This is impractical, however, due to the strong nonlinearity inherent in the vertical structure of clouds[4] and to the marked inconsistencies in cloud diurnal cycles over the land from the climate models (see Figure S9, S10 and Eq. 9). To avoid these problems, we model $\overline{R}$ by simply using a minimalist model for the earth's radiation budget and assuming one layer of atmosphere comprised of a fraction $f$ of clouds and a fraction $1$-$f$ of greenhouse gases (see Methods). The modulations in solar radiation by cloud albedo and shortwave absorption, which are critical to correctly estimate the CCP kernel, are modeled using the well-established empirical functions of cloud water path and solar zenith angle[23] (see Methods). While a single layer model only approximates the longwave radiation components, this assumption is amply justified by the fact that the longwave radiation has far less impacts on the CCP kernel (as also pointed out in previous analyses[2]), because the diurnal amplitude of solar radiation reaching the tropopause are much larger than those of longwave radiation from the earth's surface (see Figure S6).

We used the ISCCP D1 data to analyze the climatology of cloud properties such as albedo and emissivity at each geographical location in each season (see Methods). These properties were then used in the analytical solutions of radiative kernels along with the climatology of cloud diurnal cycle (see Figure 1) to obtain the geographical distribution of the CCP kernel (Figure S4). The results show that a perturbation of CCP leads to an increase of the net radiation at tropopause over the land and to a decrease over the ocean, thus changing the sign of the CCP kernels. These changing patterns are related to the characteristics of cloud cycle as shown in Figure 1, with early morning and late afternoon cloud peaks over the ocean and land, respectively. Consequently, the delay of the timing of the maximum cloud amount in the morning over the ocean causes more solar radiation shading, while the delay in the afternoon over the land implies less reflection. Besides its dependence on the timing of cloud peaks, the CCP kernel also depends on the amplitude of cloud cycle and the intensity of solar radiation (see, e.g., Eq. 9 in Methods). As a result, the larger cloud cycle amplitude over the land (see Figure 1) allows the corresponding radiative kernels to have higher absolute values than those over the ocean. The stronger solar radiation in warm seasons also contributes to reinforce the effects of the cloud diurnal cycle, causing higher absolute values of the radiative kernels (see Figure S4).

Finally, the radiative impacts ($\Delta \overline{R}$) due to the CCP shift are calculated as the multiplication of the radiative kernel ($\partial \overline{R} / \partial t_c$) and the phase shift ($\Delta t_c$). As shown in Figure 2 and Figure S5, the probability distributions of $\Delta \overline{R}$ over the land and the ocean have near-zero modes with long right tails, indicating a mean positive feedback induced by CCP shifts. Interestingly, in spite of the contrasting shifts of the CCP kernels over the oceans (shifting earlier) and over the land (shifting later), they end up having the same warming effect by reflecting less solar radiation. Geographically, the East Pacific and Atlantic oceans contribute strongly to the positive CCP feedback, while northern Asia and Australia tend to partially reduce theses radiative effects. Even though it has large CCP shifts, the West Pacific was found to have low radiative impacts due to its weak cloud cycles and small radiative kernels. Globally, the average radiative impacts ($\Delta \overline{R}$) between 60S and 60N due to the CCP shift from the colder to the warmer period in a 13-year span were found to be 0.3 W m$^{-2}$.

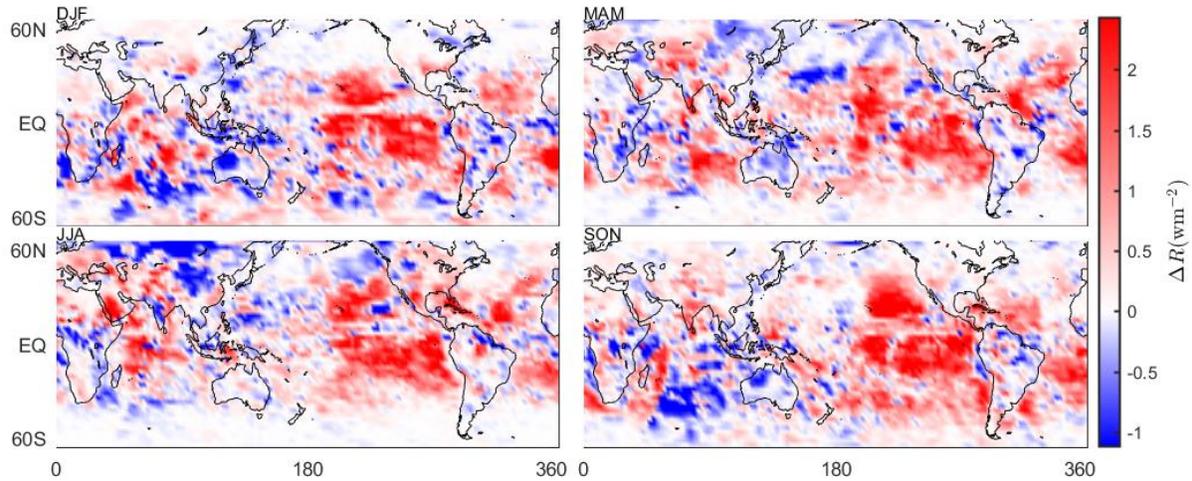

**Figure 2 / Geographical distributions of radiative impacts ($\Delta \overline{R}$) in four seasons from ISCCP records.** The radiative impacts ($\Delta \overline{R}$) is the multiplication of the radiative kernel ($\partial \overline{R} / \partial t_c$) and the CCP shift ($\Delta t_c$) from colder period (1984-1996) to warmer period (1997-2009). In the map, red color indicates $\Delta \overline{R}$ >0 and a positive CCP feedback; blue color infers $\Delta \overline{R}$ <0 and a negative CCP feedback.

It is important to bear in mind that the radiative impacts just discussed refer to decadal timescales and can be quite different from the long-term ones[24]. Moreover, they are affected by uncertainties related to the artifacts in the ISCCP records as well as to other factors besides greenhouse gas emissions (e.g. the natural internal forcing[25]). To corroborate our findings from the satellite data, we performed a similar analysis on the outputs of nine GCMs (Table S1) participating in the Fifth Phase of the Coupled Model Inter-comparison Project (CMIP5). Consistently with the ISCCP long-term climatology (see Figure S8), the mean cloud amount ($f_m$) during 1986-2005 was found to be smaller over the land and larger over the ocean. On the other hand, the amplitude and phase ($f_a$ and $t_c$) appeared to be significantly underestimated compared to well-known observations data[16,17] and showed inconsistencies among each other (see Figure S9 and Figure S10). For these reasons, in what follows we only discuss in detail the results over the oceans. These results show that, in most GCMs, the CCP tends to shift earlier (especially in GFDL-CM3, IPSL-CM5A, and HadGEM2-ES), corroborating our findings from the ISCCP records (Figure S11). These shifts are also consistent with the analyses of 'cfSites' data that show how marine clouds tend to decrease faster in the morning than in the afternoon[13], thus redistributing the clouds over the day. For each GCM, we also calculated the corresponding CCP kernels (Figure S12), using the data from historical experiments (1986-2005). Multiplying these kernels by the CCP shifts, we finally obtained the corresponding radiative impacts, which were found to be non-negligible, although affected by a large inter-model spread (see Figure 3, e.g. radiative impacts over the ocean in GFDL-CM3 can reach 0.4 W m$^{-2}$ which is 9% of the radiative forcing for the RCP45 experiment). It is reasonable to expect that the overall radiative impacts would have been even larger were the climate models able to capture the strong cloud diurnal cycles over the land[16,26].

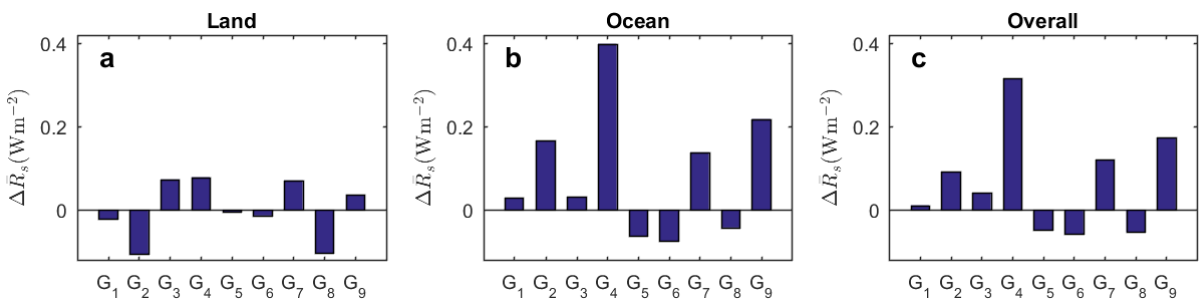

**Figure 3 / Mean shortwave radiative impacts due to CCP shifts in GCMs.** The mean radiative impacts ($\Delta \bar{R}_s$) over the land (**a**), the ocean (**b**), and the globe (**c**) within 60S-60N are due to the CCP shifts ($\Delta t_c$) from current period (1986-2005) in "historical" experiment to future period (2081-2100) in "RCP45" experiment. $G_1$ … $G_9$ refer to different GCMs listed in Table S1.

The physical mechanisms underlying these CCP shifting patterns in both ISCCP records and GCM outputs may be understood considering the low-level boundary-layer clouds. In response to warming, the atmospheric air is heated non-uniformly at different altitudes, increasing the lower-tropospheric potential temperature lapse rate[7,18]. Over land, the impacts of this lapse-rate change on cloud dynamics may be predicted by a simplified atmospheric boundary-layer (ABL) model (see Methods), previously used to simulate the timing of convective cloud formation[27,28]. The results show how the lapse-rate change slows down the growth of the ABL, thus postponing the crossing of the lifting condensation level (LCL) and resulting in a CCP shift to later times (see Figure 4). Over the ocean, the picture is more complicated, thus preventing a simple mechanistic explanation similar to the one obtained from the ABL model over land. In general, the typical marine cloud cycle (peaking at dawn and breaking up as the sun heats the cloud top[9]) is influenced by radiation-convection interaction, cloud versus cloud-free radiation differences, surface thermal variability, and life cycle of cloud system[16,29]. Moreover, stratocumulus-topped boundary layers can sustain for days over the ocean before their collapse[30]. This notwithstanding, the results are in agreement with other observations showing that, in response to the global warming, the low-level clouds generally tend to decrease faster in the morning than in the afternoon[13], consequently redistributing the clouds over the day.

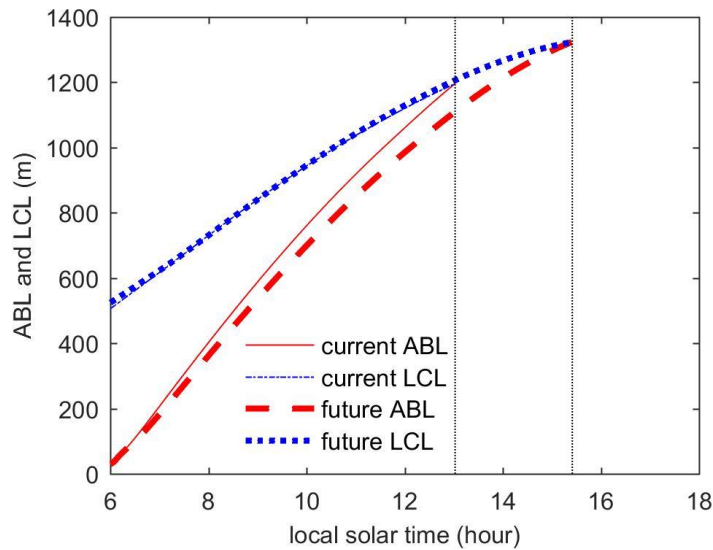

**Figure 4 / Timing of convective cloud formation in response to the global warming over the land.** The boundary-layer dynamics are simulated by a mixed-layer model (see Methods) with early morning sounding profiles from current and future climate scenarios (see details in Figure S14). Convective clouds over land are formed when the ABL crosses the LCL (vertical dash lines). The LCL crossing is delayed in future climate scenarios with increasing potential temperature lapse rate under well-watered conditions and can be further delayed under water-stressed conditions.

In summary, we have used the ISCCP satellite data and GCM outputs to quantify the changes in the cloud diurnal cycle going from colder to warmer periods and we have interpreted them in terms of radiative impacts on the global energy budget. While the timing of maximum cloud amount in warm period tends to be earlier in the morning over the ocean and slightly later in the afternoon over the land from ISCCP cloud records, both patterns seem to perform the same function of reflecting less solar radiation thus creating stronger positive feedbacks to the climate system. GCM outputs showed similar, non-negligible CCP shifting trends over the ocean, while at the same time revealing considerable inconsistencies over the land. It is hoped that the relatively large radiative impacts of the CCP shift found in this study will draw more attention to this important topic. In particular, their strong associated uncertainty should call for renewed efforts in improving cloud schemes in relation to their representation of the cloud diurnal cycle.

**Methods**

**Cloud Diurnal Cycle.** The ISCCP (http://isccp.giss.nasa.gov/) and nine GCMs listed in Table S1 participating CMIP5 (http://cmip-pcmdi.llnl.gov/cmip5/) provide long-term global cloud-fraction observation and simulation records with temporal resolution of 3 hours[15]. To find the climatology of cloud diurnal cycle, we averaged the cloud fractions at each grid box for each season over 1984-2009 for ISCCP records and over 1986-2005 for GCM "historical" experiment outputs to form a series of mean cloud fractions in a 24-hour span. This series was then decomposed into a Fourier series as

$$f(t) = f_m + f_a \cos[w(t-t_c)] + \sum_{n=2}^{N} f_n \cos[nw(t-t_n)], \tag{1}$$

where $t$ has been transformed into local solar time, $f_m$ is mean cloud fraction, $f_a$ and $t_c$ are the amplitude and phase of the first harmonic of the diurnal cycle, and $w = 2\pi/\tau$ is the angular frequency, in which $\tau$ is the length of one period of the diurnal cycle (i.e., 24 hours). The first two terms are expected to capture the main characteristics of the cloud diurnal cycle[9,31]. The mean, amplitude and phase of the cloud diurnal cycles are shown in Figure 1 and Figure S1 for ISCCP cloud records and in Figure S8, S9, and S10 for GCM simulation records. For CCP shifts in ISCCP cloud records, the same process was repeated to calculate the CCP over two periods

1984-1996 and 1997-2009. Their differences ($\Delta t_c$) are defined as the CCP shifts (Figure S2). For GCM simulations, the CCP shifts are calculated as the phase differences between "historical" (1986-2005) experiment "r1i1p1" ensemble and "RCP45" (2081-2100) experiment "r1i1p1" ensemble.

**Minimalist model for cloud cycle phase kernel.** We derived the radiative kernel based on a minimalist radiation balance model. As illustrated in Figure S7, the clear atmosphere below the tropopause is simplified as a layer of greenhouse gas, which is transparent to the shortwave radiation but semi-transparent to the longwave radiation. The layer of clouds is semi-transparent to the shortwave radiation but has high values of emissivity for longwave radiation. The net shortwave radiation at the tropopause after multiple absorptions and reflections is modeled as,

$$R_s(t) = [1-f(t)][1-\alpha_s]S(t) + f(t)[1-\alpha_c]S(t) - f(t)S(t)[1-\alpha_c-a_c]^2\alpha_s + ...$$
$$= [1-f(t)][1-\alpha_s]S(t) + f(t)[1-\alpha_c]S(t) - f(t)S(t)[1-\alpha_c-a_c]^2 \frac{\alpha_s}{1-\alpha_s\alpha_c}, \quad (2)$$

where $\alpha_c$ and $a_c$ are cloud albedo and absorptivity, $\alpha_s$ is surface albedo, and $S$ is the solar radiation reaching the tropopause. Accordingly, the net longwave radiation is

$$R_l(t) = -[1-f(t)]\left[\sigma(1-\varepsilon_g)T_s^4(t) + \sigma\varepsilon_g T_g^4(t)\right] - f(t)\left[\sigma(1-\varepsilon_c)T_s^4(t) + \sigma\varepsilon_c T_c^4(t)\right], \quad (3)$$

where $\sigma$ is Stefan–Boltzmann constant, $\varepsilon_g$ is bulk longwave emissivity of the simplified greenhouse gas layer in the clear sky, $\varepsilon_c$ is the bulk longwave emissivity of the cloudy atmosphere, and $T_c$, $T_g$, and $T_s$ are the temperature of the cloudy layer, the greenhouse gas layer, and the earth surface. The energy in the greenhouse gas layer and cloudy layer is assumed to be in equilibrium,

$$2\sigma\varepsilon_g T_g^4(t) = \sigma\varepsilon_g T_s^4(t), \tag{4}$$

and,

$$\begin{aligned}2\sigma\varepsilon_c T_c^4(t) &= \sigma\varepsilon_c T_s^4(t) + S(t)a_c + S(t)[1-\alpha_c - a_c]\alpha_s a_c + ... \\ &= \sigma\varepsilon_c T_s^4(t) + S(t)a_c + S(t)[1-\alpha_c - a_c]\frac{\alpha_s a_c}{1-\alpha_s \alpha_c}.\end{aligned} \tag{5}$$

These equations connect $T_c$ and $T_g$ to $T_s$. The net radiation at the tropopause $R(t)$ can be calculated as,

$$R(t) = R_s(t) + R_l(t) = A(t)f + B(t), \tag{6}$$

where

$$\begin{aligned}A(t) &= -S(t)[\alpha_c - \alpha_s] - S(t)[1-\alpha_c - a_c]^2 \frac{\alpha_s}{1-\alpha_s \alpha_c} \\ &+ \sigma[\varepsilon_c - \varepsilon_g]T_s^4(t) + \sigma\varepsilon_g T_g^4(t) - \sigma\varepsilon_c T_c^4(t),\end{aligned} \tag{7}$$

and

$$B(t) = S(t)[1-\alpha_s] - \sigma[1-\varepsilon_g]T_s^4(t) - \sigma\varepsilon_g T_g^4(t). \tag{8}$$

These equations succinctly explain the control of the cloud diurnal cycle on the earth's energy balance. The average net radiation over one diurnal period is $\bar{R} = 1/\tau \int_0^\tau R(t)dt$ and its derivative with respect to $t_c$ is

$$\frac{\partial \bar{R}}{\partial t_c} = \frac{1}{\tau}\int_0^\tau \frac{\partial R}{\partial t_c}dt = wf_a \frac{1}{\tau}\int_0^\tau A(t)\sin[w(t-t_c)]dt, \tag{9}$$

which is the radiative kernel of the CCP. Specifically, CCP shortwave radiative kernel is the derivative of the average net shortwave radiation $\bar{R}_s$ with respective to $t_c$,

$$\frac{\partial \bar{R}_s}{\partial t_c} = wf_a \frac{1}{\tau}\int_0^\tau A_s(t)\sin[w(t-t_c)]dt, \tag{10}$$

where

$$A_s(t) = -S(t)[\alpha_c - \alpha_s] - S(t)[1 - \alpha_c - a_c]^2 \frac{\alpha_s}{1 - \alpha_s \alpha_c}. \tag{11}$$

To obtain the parameters of CCP radiative kernel for ISCCP cloud records, we computed $S(t)$, $T_s(t)$, $\alpha_s$, $\alpha_c$, $a_c$, $\varepsilon_c$ and $\varepsilon_g$, for each geographical location and separated them for each season. The diurnal variation of solar radiation reaching the top of the atmosphere is a function of latitude, local time, and day number of the year, and 75% of this solar radiation is assumed to reach the tropopause $S(t)$[32]. $T_s(t)$ is directly obtained from ISCCP by averaging the satellite data sets to form a series of surface temperature in a 24-hour span in the same manner as the derivation of cloud diurnal cycle climatology. $\alpha_s$ is also obtained from ISCCP D1 data sets as the average reflectance of the solar radiation during the daytime. The solar zenith angle and the mean in-cloud liquid water path from ISCCP D1 data are used to calculate the $\alpha_c$ and $a_c$ for each grid box[23]. The cloud is assumed to be blackbody ($\varepsilon_c = 1$) as the water drops are efficient absorbers/emitters of longwave radiation[23,33]. To estimate $\varepsilon_g$, we consider the atmosphere comprised of greenhouse gases and clouds with mean cloud fraction 0.63 as derived from ISCCP data records. Emissivity of greenhouse gas of 0.41 combined with the cloud emissivity of unity produces an areal average atmospheric emissivity of 0.78, consistently with the estimation in simplified energy balance models[34].

The same $S(t)$ and $\alpha_s$ used to compute the ISCCP radiative kernels are also used for the GCM radiative kernels, due to the small differences in surface albedo and solar radiation reaching the tropopause in the "historical" simulations. Total liquid water path (clwvi) from GCMs is divided by the cloud fraction to calculate the average in-cloud liquid water path, which is then used to calculate the $\alpha_c$ and $a_c$ in the same manner as these in the ISCCP radiative kernels.

**Atmospheric Boundary Layer Model.** We model the dynamics of atmospheric boundary layer (ABL) with a simplified zero-dimensional mixed-layer model[35-38]. The ABL is assumed to be a well-mixed slab of the height $h$ with constant potential temperature ($\theta$) and specific humidity ($q$) profiles. The governing equations for the $\theta$ and $q$ are[37,39],

$$\rho c_p h \frac{d\theta}{dt} = H(t) + \rho c_p [\theta_f(h) - \theta] \frac{dh}{dt}, \qquad (12)$$

and

$$\rho h \frac{dq}{dt} = E(t) + \rho c_p [q_f(h) - q] \frac{dh}{dt}, \qquad (13)$$

where $\rho$ is air density, $c_p$ is the heat capacity at constant pressure, the subscript $f$ refers to the values in the free atmosphere, and surface sensible heat flux and evaporative flux ($H$ and $E$) are partitioned from the net available energy ($Q$),

$$Q(t) = \lambda E(t) + H(t), \qquad (14)$$

where $\lambda$ is the specific heat of vaporization. The energy fluxes can be modeled as[38],

$$H(t) = g_h c_p \rho [\theta_s(t) - \theta(t)], \qquad (15)$$

and

$$E(t) = g_e \rho [q_s(t) - q(t)], \qquad (16)$$

where subscript $s$ refers to the values near the surface, and the conductances to evaporative and sensible heat fluxes ($g_e$ and $g_h$) are modeled by coupling the ABL to a soil-plant system[40], in which the parameters are set for typical woody plants and loam soil[41]. To close the equations, the entrainment flux at the top of the ABL is assumed to be proportional to the surface buoyancy flux[33,42],

$$\left[\theta_{vf}(h)-\theta_v\right]\frac{dh}{dt}=-(\overline{w'\theta'_v})_h=\beta(\overline{w'\theta'_v})_s, \tag{17}$$

where the subscript $h$ refers to the values at the boundary-layer top, $\overline{w'\theta'_v}$ is buoyancy flux, and $\theta_v$ is the virtual potential temperature, which is the theoretical potential temperature of dry air with the same density as the original moist air[43],

$$\theta_v=\theta\left[1+\delta q-q_L\right], \tag{18}$$

where $\delta$ is approximately 0.61, and $q_L$ is liquid water content (equals zero in cloud-free atmosphere). Eq. (12)-(18) along with Clausius-Clapeyron equation form a closed nonlinear system for the ABL. The lifting condensation level (LCL) can be determined as the location where adiabatically lifted air parcels just become saturated,

$$q=\varepsilon\frac{e_s(T_{LCL})}{P_{LCL}}, \tag{19}$$

where $\varepsilon$ is the ratio of gas constants of dry air and water vapor, $e_s(\cdot)$ is the saturation water vapor pressure function, and $T_{LCL}$ and $P_{LCL}$ are the temperature and pressure at the LCL. Since the air parcels are dry adiabatically lifted, $T_{LCL}$ and $P_{LCL}$ follow,

$$T_{LCL}=\theta\left(\frac{P_{LCL}}{P_0}\right)^{R/c_p}, \tag{20}$$

where $P_0$ is surface air pressure, and $R$ is gas constant of moist air.

**Acknowledgments**

We gratefully acknowledge useful discussion with Minghan Zhang. We also acknowledge support from the USDA Agricultural Research Service cooperative agreement 58-6408-3-027; and National Science Foundation (NSF) grants EAR-1331846, EAR-1316258, FESD EAR-1338694 and the Duke WISeNet Grant DGE-1068871. The ISCCP satellite records were obtained from NASA Atmospheric Science Data Center (http://isccp.giss.nasa.gov/). The GISS analysis of global surface temperature change data were acquired from Goddard Institute for Space Studies (http://data.giss.nasa.gov/gistemp/). GCM outputs were downloaded from Coupled Model Inter-comparison Project website (http://cmip-pcmdi.llnl.gov/). We thank the editor and two anonymous reviewers for useful suggestions and encouragement. Models used in the paper are available upon request.

**Author Contributions**

A.P. and J.Y. conceived and designed the study. J.Y. wrote an initial draft of the paper, to which both authors contributed edits throughout.

**Additional information**

Correspondence and requests for materials should be addressed to A.P.

**Competing financial interests**

The authors declare no competing financial interests

# Supplementary Information for

# **Response and Feedback of Cloud Diurnal Cycle to Rising Temperatures**

Jun Yin[1], Amilcare Porporato[1]*

[1]Department of Civil and Environmental Engineering, Duke University, Durham, North

Carolina, USA.

*Correspondence to: amilcare.porporato@duke.edu

## 1. Supplementary figures from ISCCP records

This section presents supplementary figures for estimating radiative impacts due to CCP shifts from the ISCCP satellite records. Figure S1 (a supplementary figure for Figure 1) shows the climatology of cloud diurnal cycle; Figure S2 and Figure S3 display the seasonal and geographical distributions of CCP shifts; Figure S4 provides the radiative kernels for CCP shifts; Figure S5 gives the statistics of radiative impacts due to CCP shifts; Figure S6 compares diurnal amplitude of longwave and shortwave radiation.

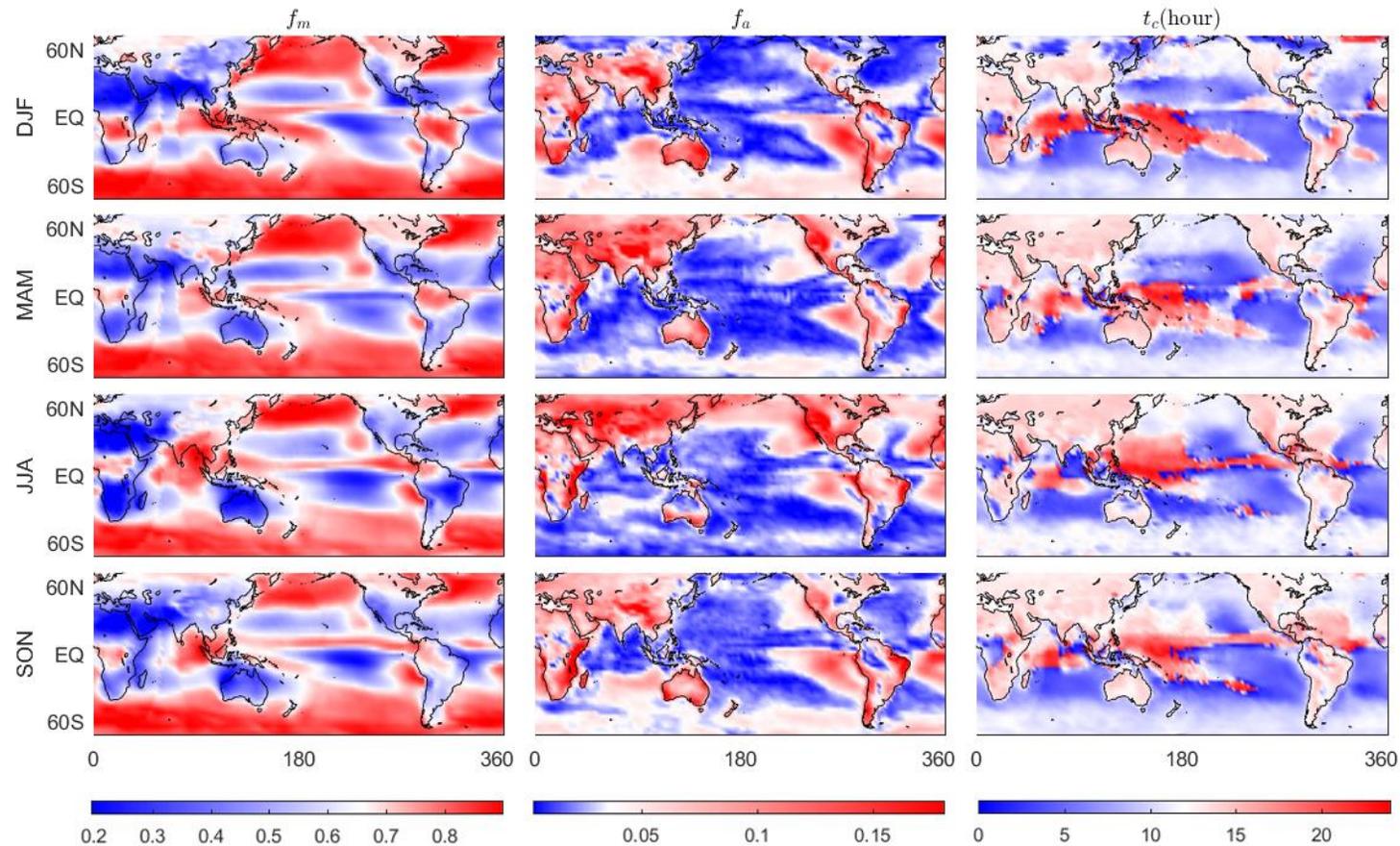

**Figure S1. Climatology of cloud diurnal cycle from ISCCP records.** (left to right) mean, amplitude, and phase of the first harmonic of the cloud diurnal cycle (top to bottom) in four seasons from ISCCP D1 data during 1984-2009. The white color represents the mean values of the corresponding variables; the blue and red indicate the values are below and above the average, respectively. The strips over the Indian Ocean are due to the lack of geostationary satellite before the late 1990s[17,20].

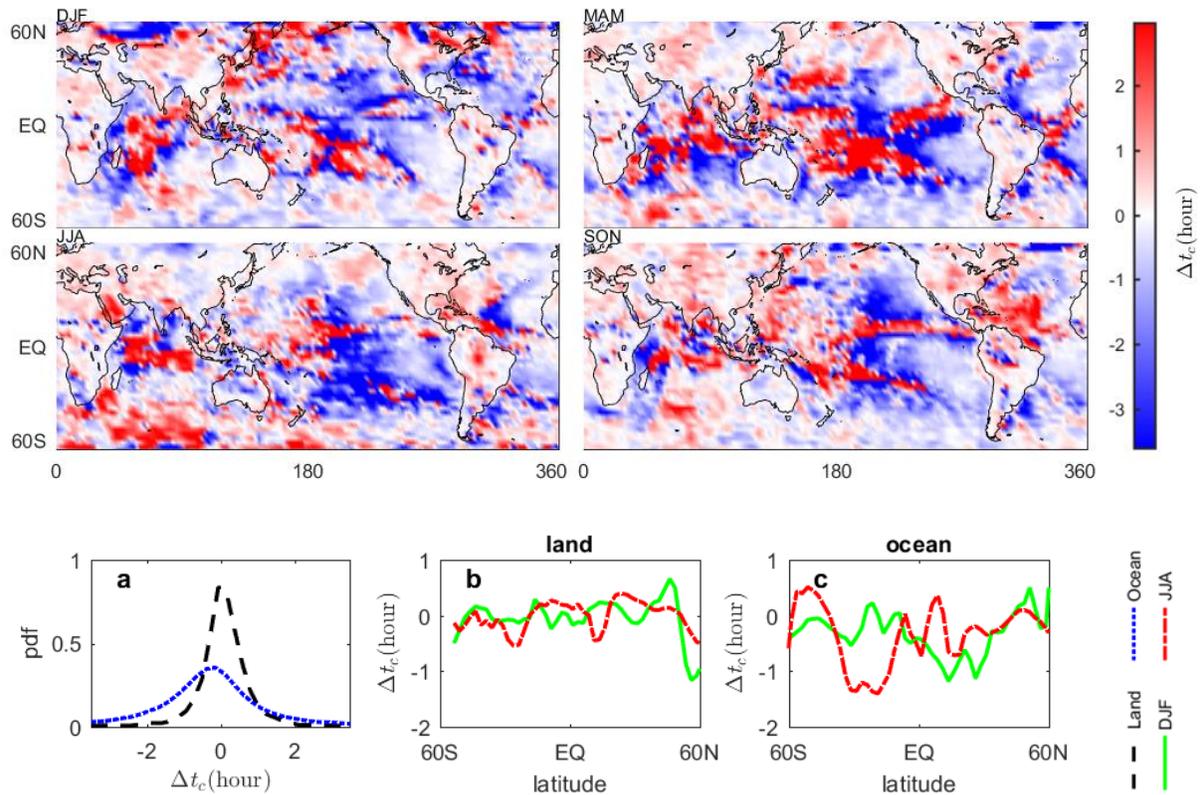

**Figure S2 / CCP shift ($\Delta t_c$) from ISCCP records.** The cloud cycle phase in relatively colder period 1984-1996 ($t_{c,c}$) and in relatively warmer period 1997-2009 ($t_{c,w}$) are computed from ISCCP D1 3-hour data sets; the phase shift is the difference between the two ($\Delta t_c = t_{c,w} - t_{c,c}$). Top panel: geographical distributions of $\Delta t_c$ in four seasons. Bottom panel: (**a**) The probability distribution function (pdf) of $\Delta t_c$ over the land (black dash) and the ocean (blue dot); zonal average of $\Delta t_c$ over (**b**) the land and (**c**) the ocean in boreal summer (red dash-dot) and winter (green solid).

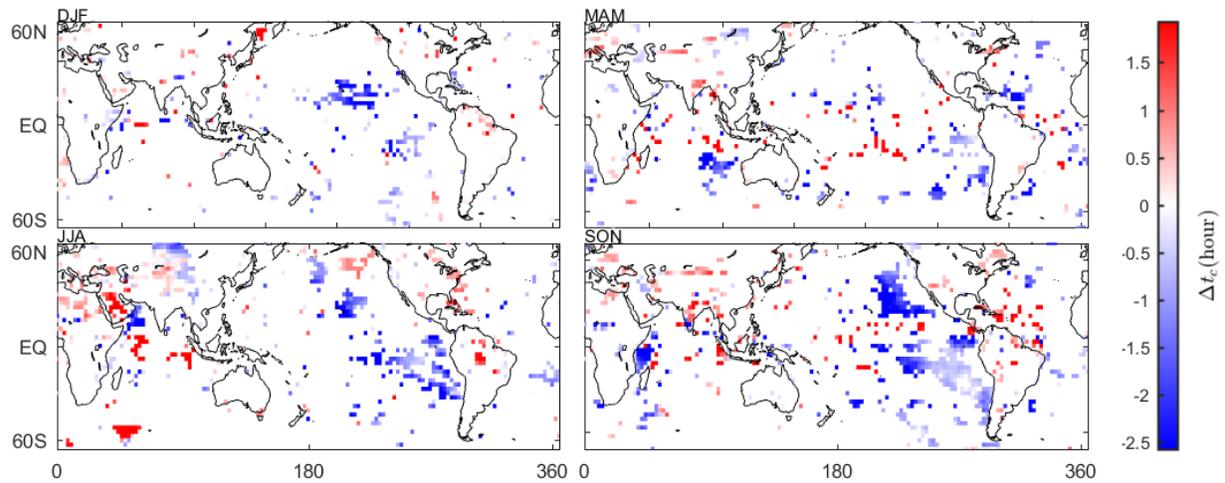

**Figure S3 / As in Figure S2 but in regions of strong correlation between CCP and surface temperature.** The correlations between time series (1984-2009) of CCP from ISCCP D1 data sets and surface temperature time series from GISS records are statistically significant in regions marked with red (positive correlation) and blue (negative correlation) colors under significance level 0.05.

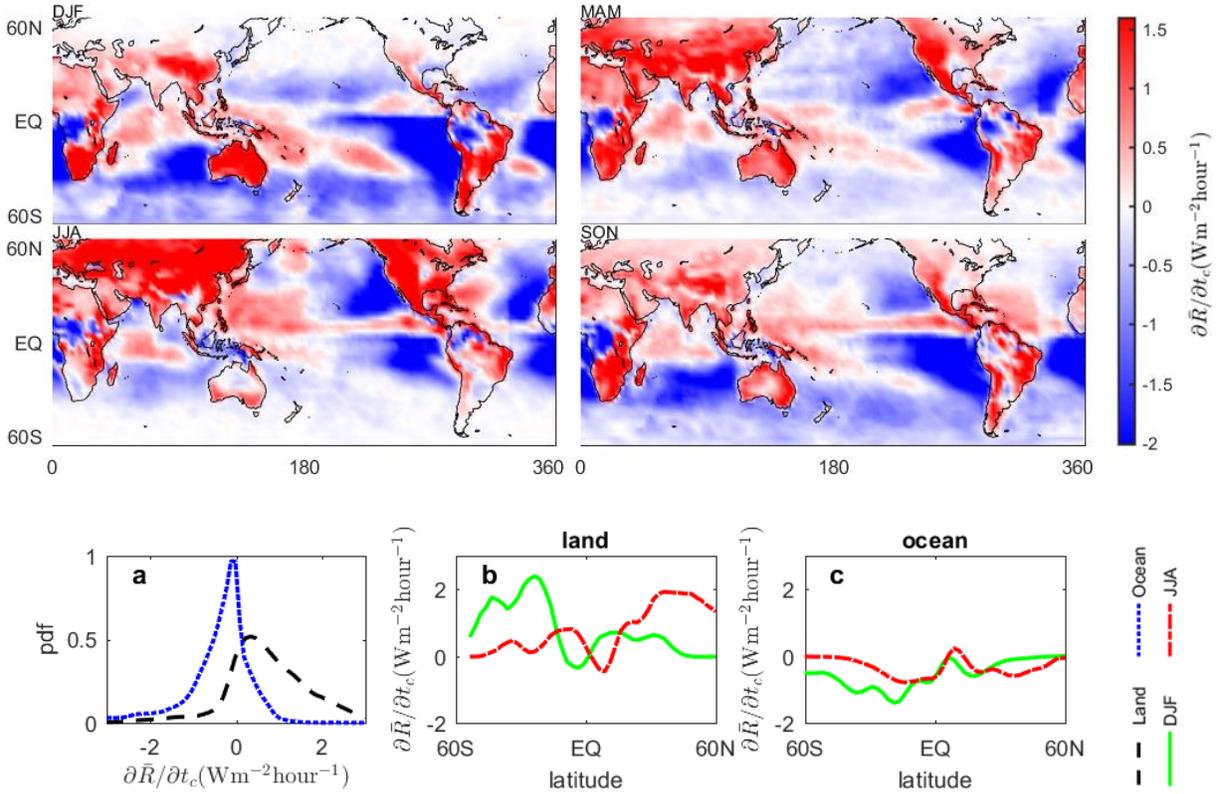

**Figure S4 / CCP kernel ($\partial \overline{R}/\partial t_c$) from ISCCP records.** The kernel is computed using the partial radiative perturbation method with cloud properties estimated from ISCCP D1 satellite records during 1984-2009 (see Methods). Top panel: geographical distributions of $\partial \overline{R}/\partial t_c$ in four seasons. Bottom panel: (**a**) The probability distribution function (pdf) of $\partial \overline{R}/\partial t_c$ over the land (black dash) and the ocean (blue dot); zonal average of $\partial \overline{R}/\partial t_c$ over (**b**) land and (**c**) ocean in boreal summer (red dash-dot) and winter (green solid).

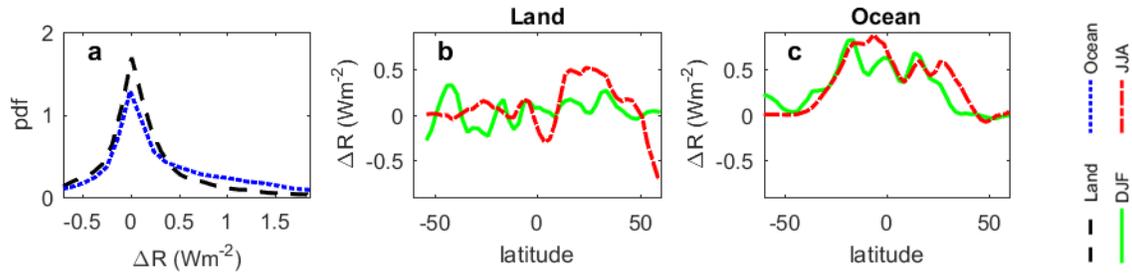

**Figure S5 / Statistics of radiative impacts due to cloud cycle phase shift ($\Delta \overline{R}$) from ISCCP records.** These radiative impacts are obtained by multiplying the radiative kernel by the cloud cycle phase shift. (**a**) The probability distribution function (pdf) of $\Delta \overline{R}$ over the land (black dash) and the ocean (blue dot); zonal average of $\Delta \overline{R}$ over (**b**) the land and (**c**) the ocean in boreal summer (red dash-dot) and winter (green solid).

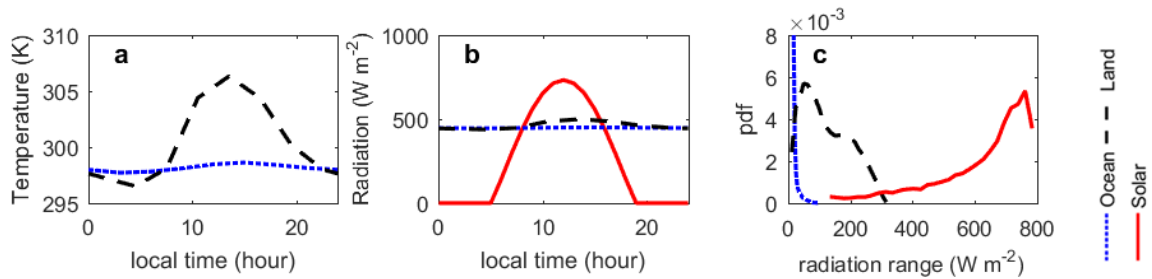

**Figure S6 / Diurnal variation of radiation and surface temperature from ISCCP records.** In (**a**), the black dash and blue dot lines show the mean diurnal variation surface temperature in boreal summer over the North Pacific Ocean (31.25N, 180E; blue dot) and East Asia (31.25N, 112.68E; black dash). In (**b**), the black dash and blue dot lines present the mean surface longwave radiation in boreal summer at the same locations of North Pacific Ocean and East Asia; The red solid line also shows the diurnal variation of solar radiation reaching the tropopause $S(t)$ at the same latitude 31.25N in boreal summer. In (**c**), the empirical probability distribution functions (pdf) of the diurnal ranges of radiation within 60S-60N are presented. The blue dot line, black dash line, and red solid line refer to the longwave radiation range over the ocean, longwave radiation range over the land, and range of solar radiation $S(t)$, respectively.

## 2. Schematic diagram for minimalist radiation balance model

Figure S7 illustrates radiation components in the minimalist radiation balance model as further explained in the Methods section.

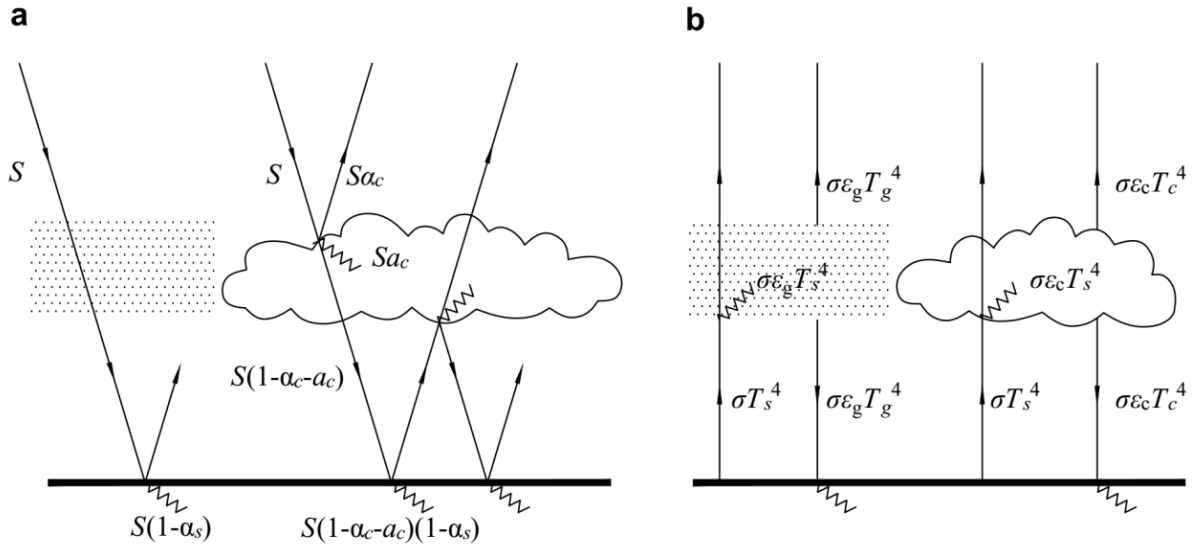

**Figure S7 / Schematic diagram of radiation components for the minimalist radiation balance model.** The dot area and cloud-shaped area represent layers of clear and cloudy atmosphere, respectively. In (**a**), the solar radiation passes through the clear atmosphere, while it is partially reflected and absorbed by the clouds with albedo $\alpha_c$ and absorptivity $a_c$. In (**b**), the clear atmosphere is assumed to be a grey body with emissivity $\varepsilon_g$, which absorbs and reemits the longwave radiation; the clouds also redistribute the longwave radiation in the same manner but with much higher emissivity $\varepsilon_c$.

# 3. Supplementary figures from GCM outputs

This section presents supplementary information for estimating radiative impacts due to CCP shifts from the GCM outputs. Table S1 lists the GCMs used in this study for assessing the CCP feedbacks. Figure S8, Figure S9, and Figure S10 show the statistics of mean, amplitude, and phase of cloud diurnal cycle from GCM 'historical' (1986-2005) experiments. Figure S11, Figure S12, and Figure S13 provide the statistics of CCP shifts, radiative kernels, and radiative impacts.

**Table S1 Climate Models used for assessing CCP feedbacks**

| No. | Acronyms | Model Institutions and References |
|---|---|---|
| 1 | CMCC-CM | The Euro-Mediterranean Center on Climate Change, Italy[44] |
| 2 | CNRM-CM5 | National Center for Meteorological Research, France[45] |
| 3 | FGOALS-g2 | LASG, Institute of Atmospheric Physics, Chinese Academy of Sciences, China; CESS, Tsinghua University, China[46] |
| 4 | GFDL-CM3 | NOAA Geophysical Fluid Dynamics Laboratory, USA[47] |
| 5 | GFDL-ESM2G | NOAA Geophysical Fluid Dynamics Laboratory, USA[48] |
| 6 | GFDL-ESM2M | NOAA Geophysical Fluid Dynamics Laboratory, USA[48] |
| 7 | HadGEM2-ES | Met Office Hadley Centre, United Kingdom[49] |
| 8 | INM-CM4 | Institute for Numerical Mathematics, Russia[50] |
| 9 | IPSL-CM5A-MR | Institute Pierre Simon Laplace, France[51] |

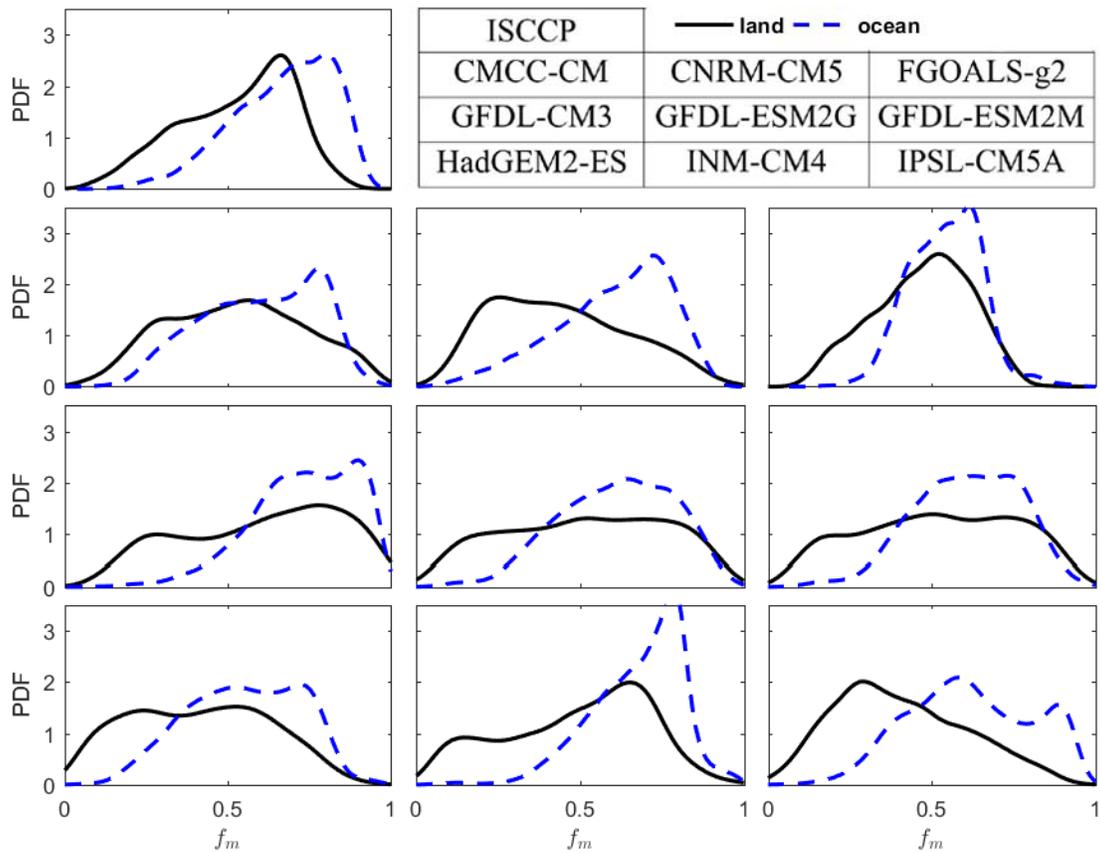

**Figure S8 / PDFs of mean cloud amount ($f_m$) over the land and ocean from GCM outputs and ISCCP records.** Overall, the mean cloud amount over land is smaller than that over the ocean from nine GCMs outputs and the ISCCP cloud records. The inset at the top right corner maps the names of the data sources onto the corresponding tiled blocks with the same relative positions. The GCMs outputs are from "historical" (1986-2005) experiment "r1i1p1" ensemble member and the ISCCP D1 records cover the period 1984-2009. The variable is sampled in all four seasons and in approximately equal-area grids between 60S-60N of latitude for both ISCCP and GCMs.

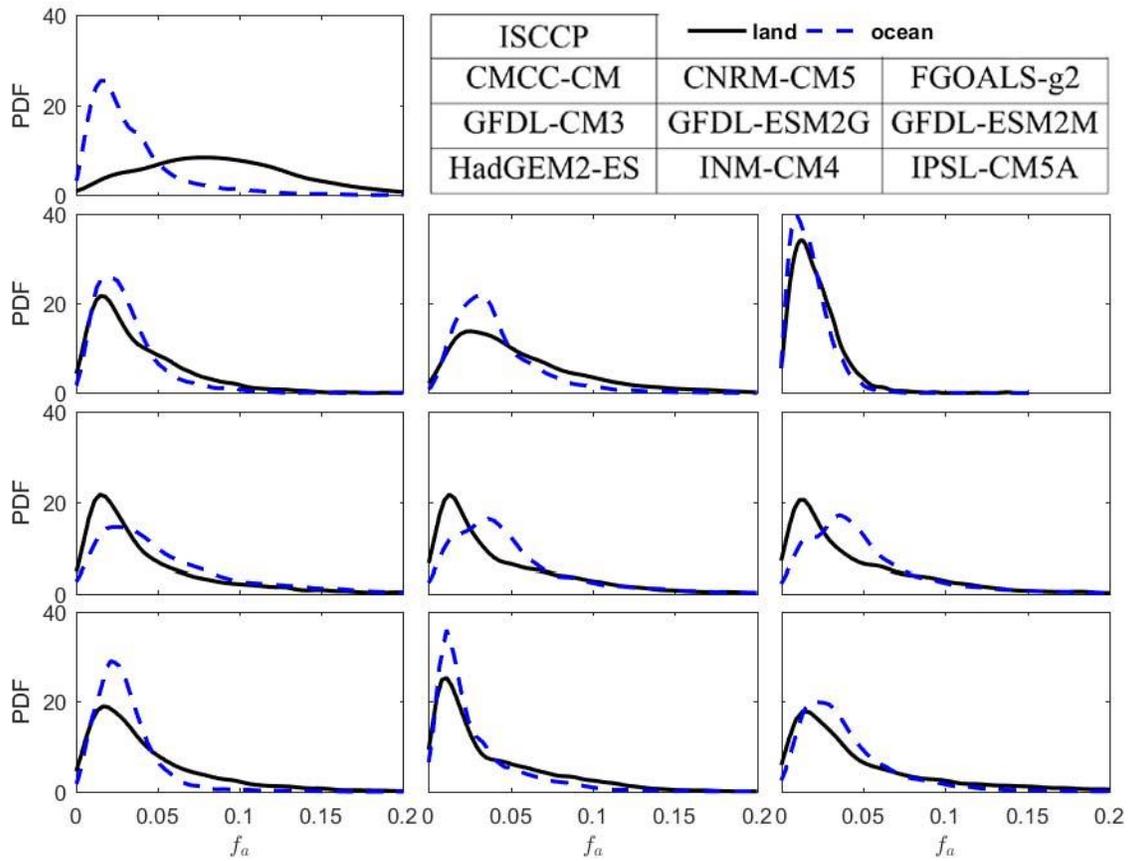

**Figure S9 / As in Figure S8 but for the amplitude of cloud diurnal cycle ($f_a$).** While it is expected that amplitude of cloud diurnal cycle is stronger over the land than over the ocean as demonstrated in the ISCCP records and many other studies[16,26,52,53], these nine GCMs tend to consistently underestimate $f_a$ over the land.

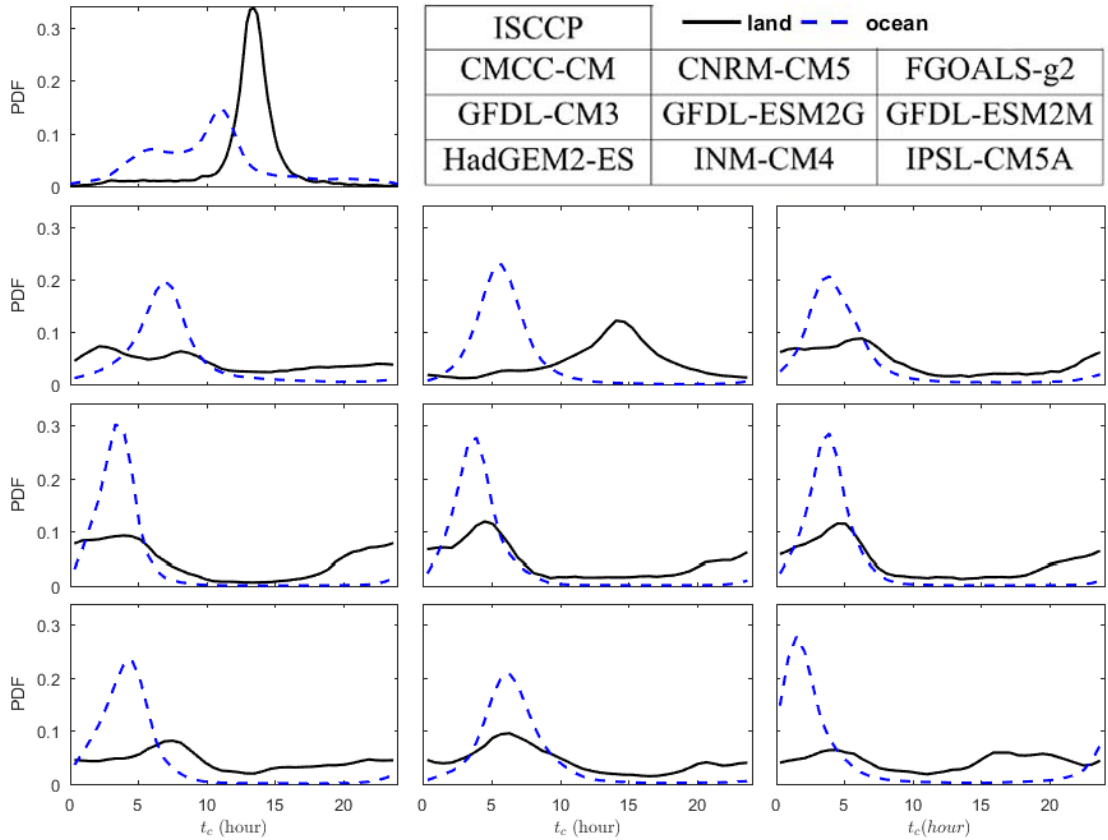

**Figure S10 / As in Figure S8 but for CCP ($t_c$).** While it is expected that cloud fraction frequently peaks in the early morning over the ocean and in the afternoon over the land as demonstrated in the ISCCP records and many other studies[16,26,52,53], these GCMs do not capture the afternoon cloud peak over the continental regions except the CNRM-CM5 model.

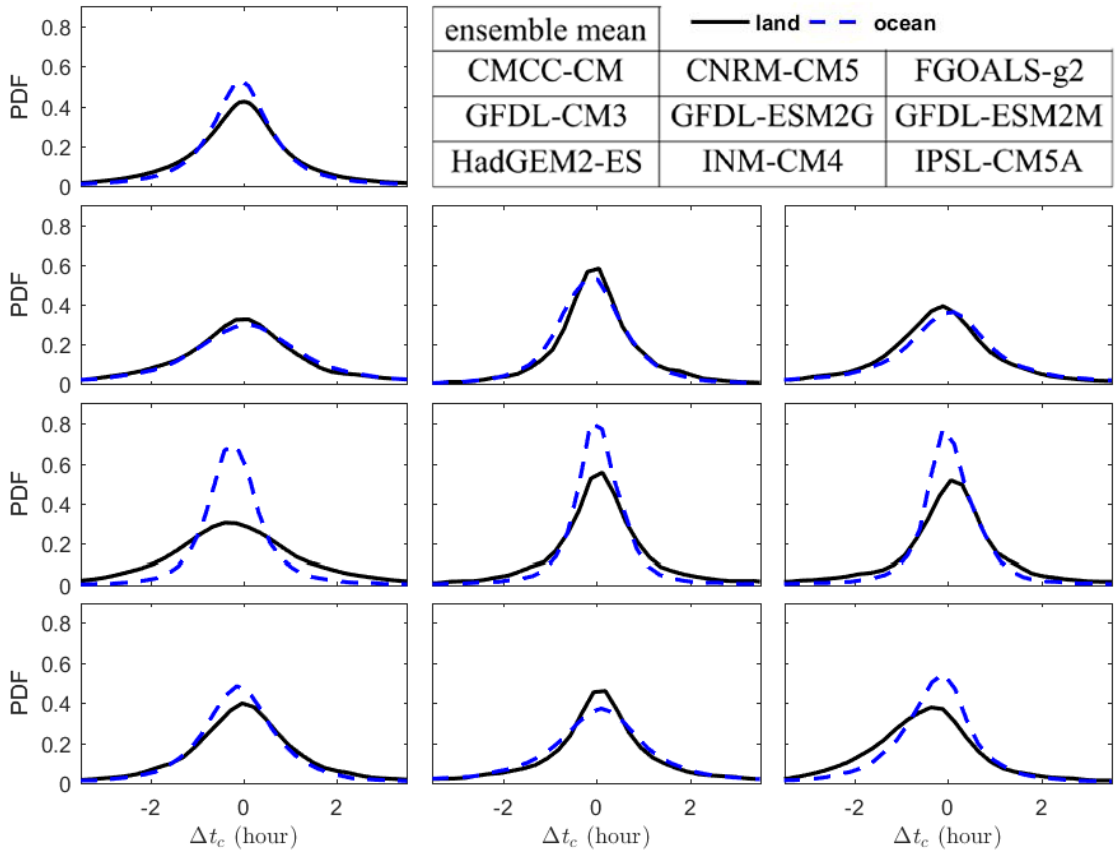

**Figure S11 / PDF of CCP shifts ($\Delta t_c$) over the land and ocean from GCM outputs.** The phase shifts are the differences between the CCP from "historical" (1986-2005) experiment "r1i1p1" ensemble and from "rcp45" (2081-2100) experiment "r1i1p1" ensemble. The inset at the top right corner maps the names of the data sources onto the corresponding tiled blocks with the same relative positions. The variable is sampled in all four seasons in approximately equal-area grids between 60S-60N of latitude.

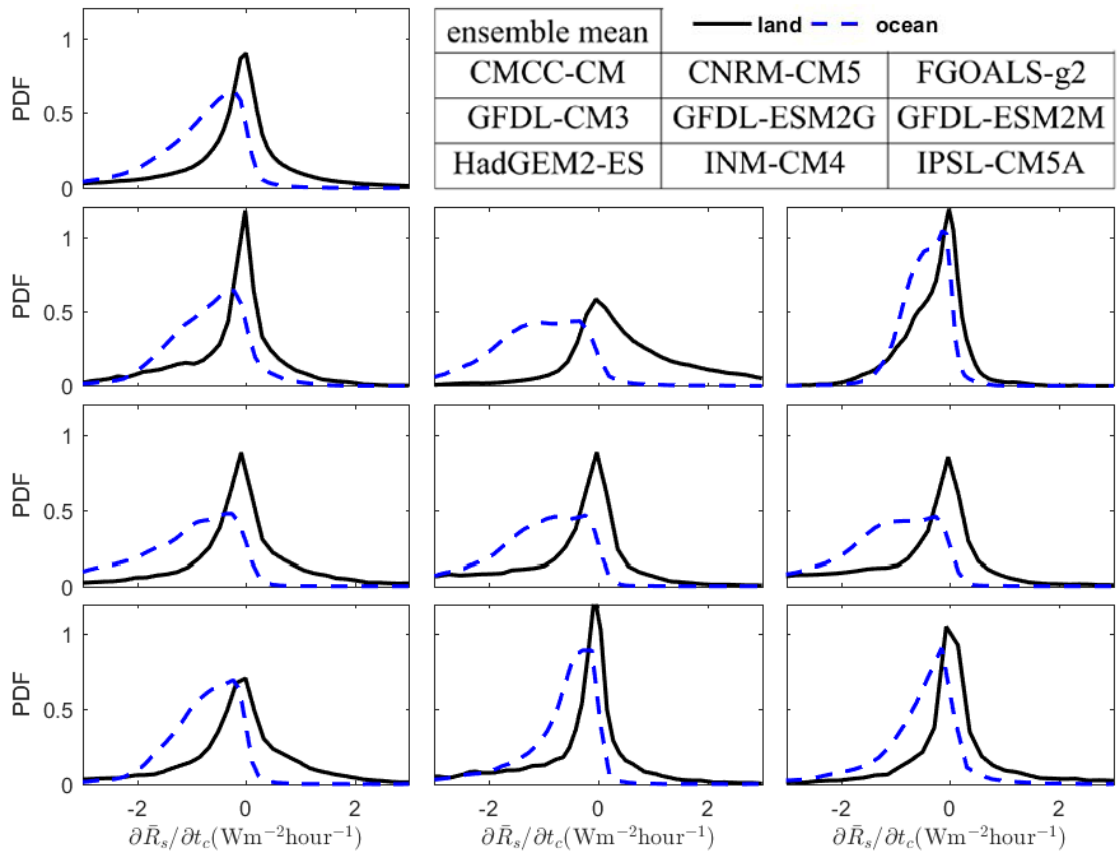

Figure S12 / As in Figure S11 but for CCP shortwave radiative kernels ($\partial \bar{R}_s / \partial t_c$).

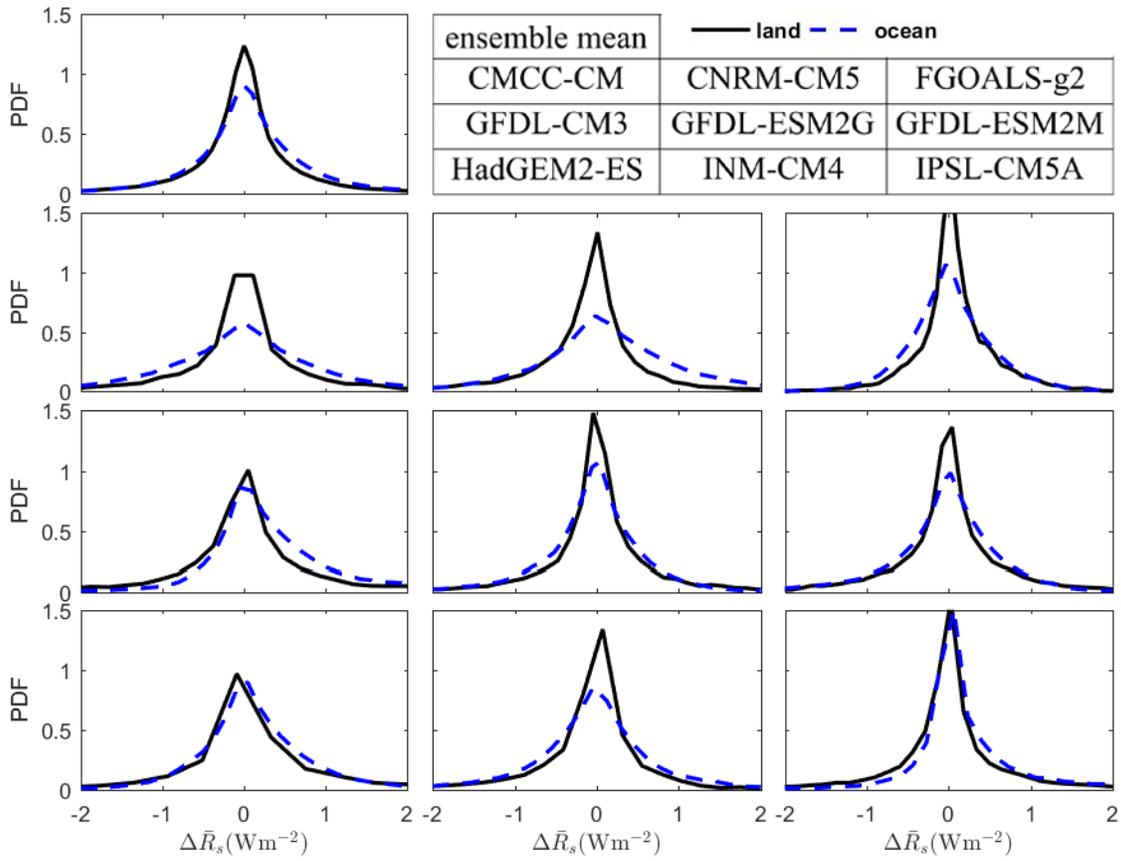

**Figure S13 / As in Figure S11 but for shortwave radiative impacts due to CCP shifts ($\Delta \bar{R}_s$).**

## 4. Morning sounding profiles used for analyzing atmospheric boundary-layer dynamics

Figure S14 presents the typical current and future morning sounding profiles used in a simplified mixed-layer model (see Methods) to simulate the atmospheric boundary-layer dynamics. Initiation of moist convection, as the first time when the atmospheric boundary layer crosses the lifting condensation level, is delayed under future climate condition due to the change of atmospheric lapse rate (see results in Figure 4). More comprehensive analyses of the initiation of moist convection in response to the changing land and atmosphere conditions (including surface hydrological controls) can be found in some recent studies[41,54].

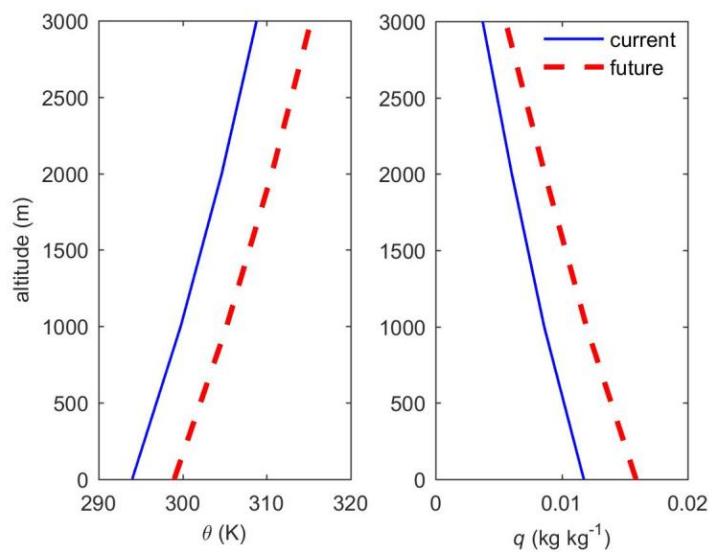

**Figure S14 / Typical morning sounding profiles under current and future climate conditions.** The current potential temperature ($\theta$) and specific humidity ($q$) profiles are typical sounding profiles in mid-latitude summer[55]. The future temperature profiles are estimated by linearly increasing 4K near the surface and 6K at 3500m due to the non-uniform change of temperature profiles[18]. The future specific humidity profiles are derived from one likely assumption that the relative humidity keeps the same with increasing temperatures[4]. The available energy is approximated by a parabolic function with midday peak values of 500 Wm$^{-2}$ in current climate condition and 510 Wm$^{-2}$ in future climate scenario to account for potential global warming effects. These morning sounding profiles and boundary conditions are used in a simplified atmospheric boundary-layer model (see Method) to simulate the timing of convective cloud formation as shown in Figure 4 for explaining potential mechanisms of CCP feedbacks.